\documentclass{osa-article}

\journal{oe}

\usepackage{color}
\usepackage{todonotes}
\usepackage{hyperref}
\hypersetup{colorlinks=true,citecolor=blue}

\begin{document}

\title{Convolutional neural network for self-mixing interferometric displacement sensing}

\author{Stéphane Barland \authormark{1,*}  and François Gustave\authormark{2}}

\address{\authormark{1}Université Côte d'Azur - CNRS, Institut de Physique de Nice, 1361 route des Lucioles, F-06560, Valbonne, France\\
\authormark{2}DOTA, ONERA, Université Paris-Saclay, F-91123, Palaiseau, France}

\email{\authormark{*}stephane.barland@univ-cotedazur.fr} 

\newcommand{\steph}[2][inline]{\todo[#1, color=blue!5!white]{\small \texttt{SB says}: #2}}

\newcommand{\ffig}[1]{Fig.~\ref{#1}}


\begin{abstract*}
	Self mixing interferometry is a well established interferometric measurement technique. In spite of the robustness and simplicity of the concept, interpreting the self-mixing signal is often complicated in practice, which is detrimental to measurement availability. Here we discuss the use of a convolutional neural network to reconstruct the displacement of a target from the self mixing signal in a semiconductor laser. The network, once trained on periodic displacement patterns, can reconstruct arbitrarily complex displacement in different alignment conditions and setups. The approach validated here is amenable to generalization to modulated schemes or even to totally different self mixing sensing tasks.
\end{abstract*}


\makeatletter
\@input{modtime}
\makeatother

\section{Introduction}

Optical interferometric measurements are routinely used in science and engineering and many schemes can be used to adapt the approach to the specific measurement to be performed. One particularly interesting and well established method is the so-called self-mixing interferometry, which consists in realizing interference between the beam reflected by a target and a reference beam \textit{inside} the laser resonator emitting the reference beam (see \textit{eg} \cite{giuliani2002laser,kane2005unlocking,donati2012developing,taimre2016laser,li2017laser} for reviews). For its simplicity and versatility, many applications have been envisioned and perhaps the most immediate is that of displacement measurement. Two limit regimes are considered \cite{donati2018overview}: that of very small displacement (much smaller than the laser wavelength) or the opposite case where the displacement takes place over a very large number of wavelengths. In the first case, information about the target displacement can be retrieved from fitting the shape of the interferometric signal. In the latter case, most of the information is obtained by counting the fringes that are observed as a sawtooth signal whose symmetry depends on the direction of the motion. Despite its apparent simplicity, this analysis is often complicated since the exact shape of the interferometric signal depends on many factors including bias current, target reflectivity, alignment conditions \cite{addy1996effects}, and modal structure of the laser which may even lead to a double-peak structure in each fringe \cite{lv2005effect}. Furthermore, on diffusive targets, speckle leads to an effective variation of the feebdack parameters and therefore a change of the signal shape in the course of the measurement. In practice, all these effects tremendously affect the \textit{availability} of self-mixing measurement setups. This has led to a number of hardware and software proposals to either improve the signal quality or the retrieval of the displacement from the interferometric signal \cite{norgia2001interferometric,zabit2010adaptive,bernal2014robust,arriaga2014speckle,usman2019detection,siddiqui2017_all}.

Computer neural networks are one of the many architectures which can be used for machine learning tasks, whereby a computer is used to infer rules from a set of data and results instead of providing results on the basis of an input and \textit{a priori} known rules. The training of a neural network leads to the formulation of kind of statistical model \cite{goodfellow2016deep}, able to predict new results on the basis of new data. These neural networks are already very widely used in everyday life and they are proving increasingly useful many areas of research and technology. In the specific context of interferometry, very few attempts exist to date. They have been used to identify and count fringes in \cite{li2019deep,reyes2019deep,kou2020fringe}. In \cite{wei2007pre} and \cite{ahmed2019self} they have been used to pre-process self mixing traces and in \cite{wang2020real} they are used as a part of a self-mixing blood pressure measurement scheme. 

In the following, we discuss the use of a convolutional neural network for the direct reconstruction of a displacement signal across many different alignment conditions. We address a particularly delicate regime which is the one of "few wavelengths" displacement. The neural network is first trained on a set of periodic data and in different alignment conditions. Its performance in reconstructing the displacement of a target from a self-mixing interferometric signal is then validated on aperiodic times series whose continuous spectrum spans more than three octaves and under different alignment conditions, not used during training. We also analyze the robustness of the reconstruction to the presence of very strong detection noise. Finally, we observe that the neural network can, without any tuning, provide a sensible reconstruction of the displacement of a target obtained on a different experimental setup based on the same operating principle. We then briefly discuss some details on the operation of such a network and some further possible uses of this approach in the context of self mixing. Thus, a reasonably simple neural network such as the one used here can become one of the tools which contribute to the robustness and high availability of self-mixing interferometric setups.

\section{Neural network design and training}

\subsection{Experimental setup}

\begin{figure}
	\centering
	\includegraphics[width=0.9\textwidth]{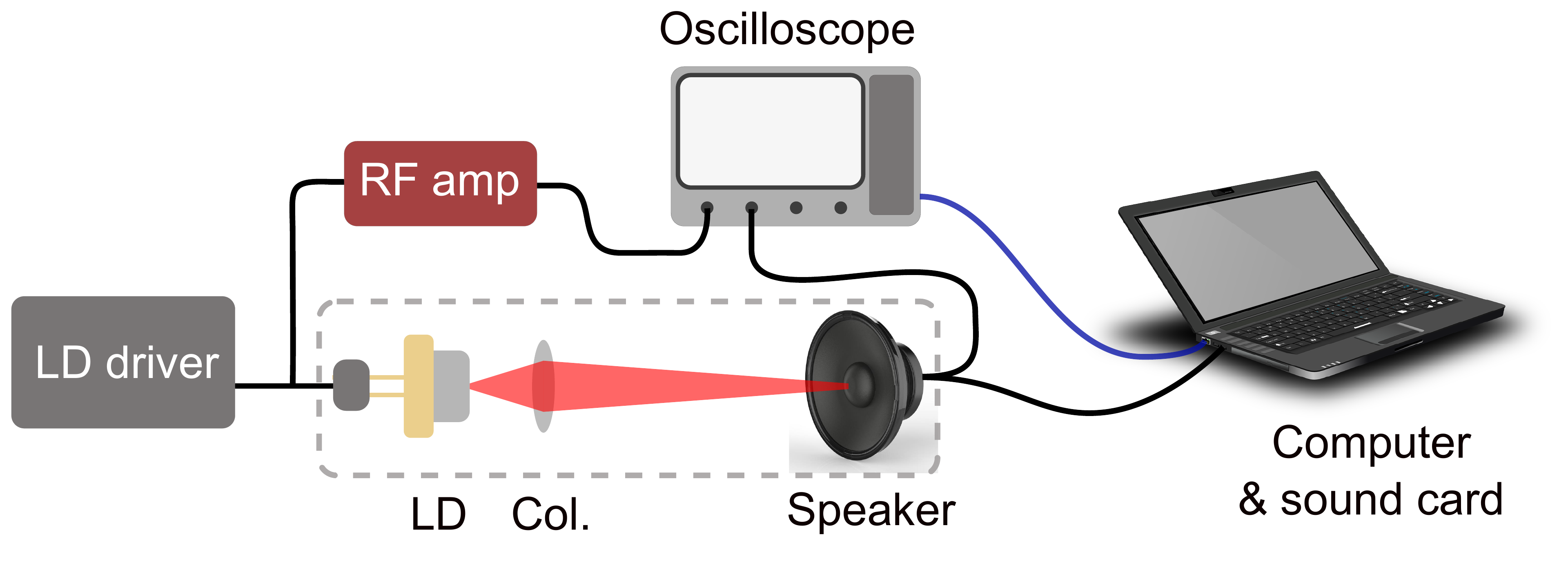}
	\caption{The self-mixing experimental setup consists of a laser diode (LD), short focal length collimator (Col.), laser diode current driver (LD driver) and voltage amplifier (RF amp).  \label{fig:setup}}
\end{figure}
The experimental arrangement, presented in \ffig{fig:setup}, consists of a single transverse mode laser emitting at $\lambda=1310~nm$ (ML725B8F) whose threshold current is about $6.5~mA$ . In all the experiments reported here the laser is driven at a constant current of 9~mA. The laser beam is focused by a high numerical aperture (NA=0.7) lens on the central region of a basic computer speaker located at about 20~cm from the laser. This speaker is put in motion via an electrical signal produced by the sound card of a computer which can easily produce many kinds of patterns at a sampling rate of 44.1~kHz. The linearity of the speaker response has been assessed over a range of frequencies from 5~Hz to 100~Hz and over the range of voltage provided by the sound card. Over this range, the speaker responds with constant 0 phase. Thus, in the range where the linearity of the displacement has been checked, one can use the voltage at the speaker as an independent measurement of the position of the target with respect to some unknown origin. From that point on, we will therefore use this voltage as a proxy for the target position. The self mixing signal is measured as a voltage at the laser electrodes, which is amplified by an AC-coupled amplifier with $10^4$ amplification factor and several MHz bandwidth. We deliberately did not optimize the self-mixing signal quality, exactly because one of our aims is to check that neural networks can help in making the measurement work even in sub optimal conditions. 

\subsection{Network setup}

The first thing to be noted is that in the "few wavelengths" range of displacement, the self-mixing signal contains no information about the absolute position. Although one can be tempted to consider that counting fringes (or some other equivalent technique) will lead to knowledge of the exact position with respect to some unknown arbitrary origin, this approach is bound to diffuse in the long term: If for some reason a fringe is missed, the measurement system has no way to recover from this error because the physics of the system does not include this information. Thus, (independently of how accurate the fringe counting is unless it is \textit{strictly perfect}), a position measurement will unavoidably loose accuracy at a rate proportional to $\sqrt{t}$ where $t$ is the measurement duration. Therefore, our aim here is to provide a measurement of the displacement within a prescribed time interval, a \textit{velocity}.

Once this is established, the setting of the architecture of the neural network is strongly influenced by the specific question one wants to address. Here we assume that the self-mixing signal is acquired at a much larger sampling rate than the Nyquist frequency of the displacement signal one wants to measure. This is a very reasonable requirement in this context since most approaches address the question by counting fringes. Here we assume that the signal is sampled at least 256 times faster than the Nyquist frequency of the displacement to be measured. Therefore, the reconstruction of the trajectory consists in inferring from 256 self-mixing signal points one single instantaneous velocity corresponding to the displacement of the target during the 256 points acquisition. Then, in terms of machine learning the problem is reduced to a "regression" problem, where some algorithm must provide a single number on the basis of the available information (a piece of time trace of length 256).

The setting therefore consists in analyzing a \textit{sequence} where temporal ordering matters and therefore a recurrent neural network can be envisioned as a suitable architecture. However, these networks are notoriously difficult to train and convolutional neural networks are known to be an easier to train and valid alternative alternative. Therefore, we build a network based on a stack of 1-dimensional convolutional layers with pooling layers between two convolutional layers. At the end of the stack, two fully connected layers convert the features identified by the convolutional layers into a single number which is the inferred velocity of the target during the measurement sequence. More details are given in appendix, table \ref{tab:network}. This global architecture was chosen from first principles of neural network design \cite{goodfellow2016deep} and the model details where then determined empirically. The network was implemented with the Keras library, which offers an excellent tradeoff in terms of complexity and versatility for our purpose \cite{chollet2015keras}.

\subsection{Network training}

Once the network architecture is chosen, the network must be trained with known data. In practice, that means providing the network a large number of pairs $[s(t_0,...,t_0+256~dt), v]$ where $s(t)$ is a self-mixing signal acquired during 256 sampling times $dt$ and $v$ is the average velocity of the target during the duration of the interval $256~dt$. One must underline that neural networks are known to be able to represent arbitrary functions provided a sufficient number of layers and cells are present in the network \cite{hornik1989multilayer}. Therefore, given enough computer time for training, a sufficiently  large network will be able to perfectly reproduce the training data it has been shown. This means that a model trained this way achieves excellent \textit{accuracy}. However, one of the key issues with self-mixing implementations is that the alignment conditions are sometimes different from one measurement to the next. Equivalently, speckle generated by the reflection of the beam on a diffusive target will lead to effective variations of the feedback strength parameter in the course of the measurement. Therefore, the network trained here must be able to adapt to these changes. This is known as the capacity of the network to \textit{generalize} the features which were learnt and identify them in unseen data. To achieve this, we train the network on a deliberately limited set of data and observe the reconstruction of the network on a very different data set, both in terms of the dynamics of the target (different displacement patterns) and in terms of the alignment of the beam on the target. The training data consists exclusively of measurements of the interferometric signal in response to periodic displacement of the target. We record self mixing signals in six different alignment conditions, in three of them a double peak is visible in each fringe. For each of these alignment conditions, we record the self mixing signal for a set of 19 frequencies evenly spaced between 10 and 100Hz. For each frequency we record 5 different amplitude signals. For each of these settings we record sinusoidal and triangular waveforms. The sampling rate of the oscilloscope is set to 250~kHz so that $dt=4\mu$s. In total the network is trained on about $1.95\times10^3$ segments of 256 time steps, each of them of duration $256 * 4\mu$s~$= 1.024$~ms. The operation regime of self-mixing sensing setups is often characterized in terms of the $C$ feeback parameter. Here, the alignment configurations we use are such that the system operates in the weak feedback regime $C<1$ (we do not observe multistability). However, we also avoid the weakest feeback regime $C<<1$ in which the interferometric signal is symmetric since it does not carry the relevant information. As is common in deep learning network training, the data is further augmented by adding noise to the training set. Here we add a delta-correlated gaussian noise on top of the measured interferometric signal. It is trained by minimizing the mean squared error between a guess it provides and the known measured displacement.

\section{Results}

After training, one will assess the performance of the neural network (also "the model") by comparing the displacement reconstructed from the interferometric signal and the voltage at the speaker's ends, used as a proxy of position. First, we check the model's prediction accuracy in known settings (periodic signals and known alignment conditions) and then in unseen settings.

\begin{figure}
	\centering
	\includegraphics[width=0.6\textwidth]{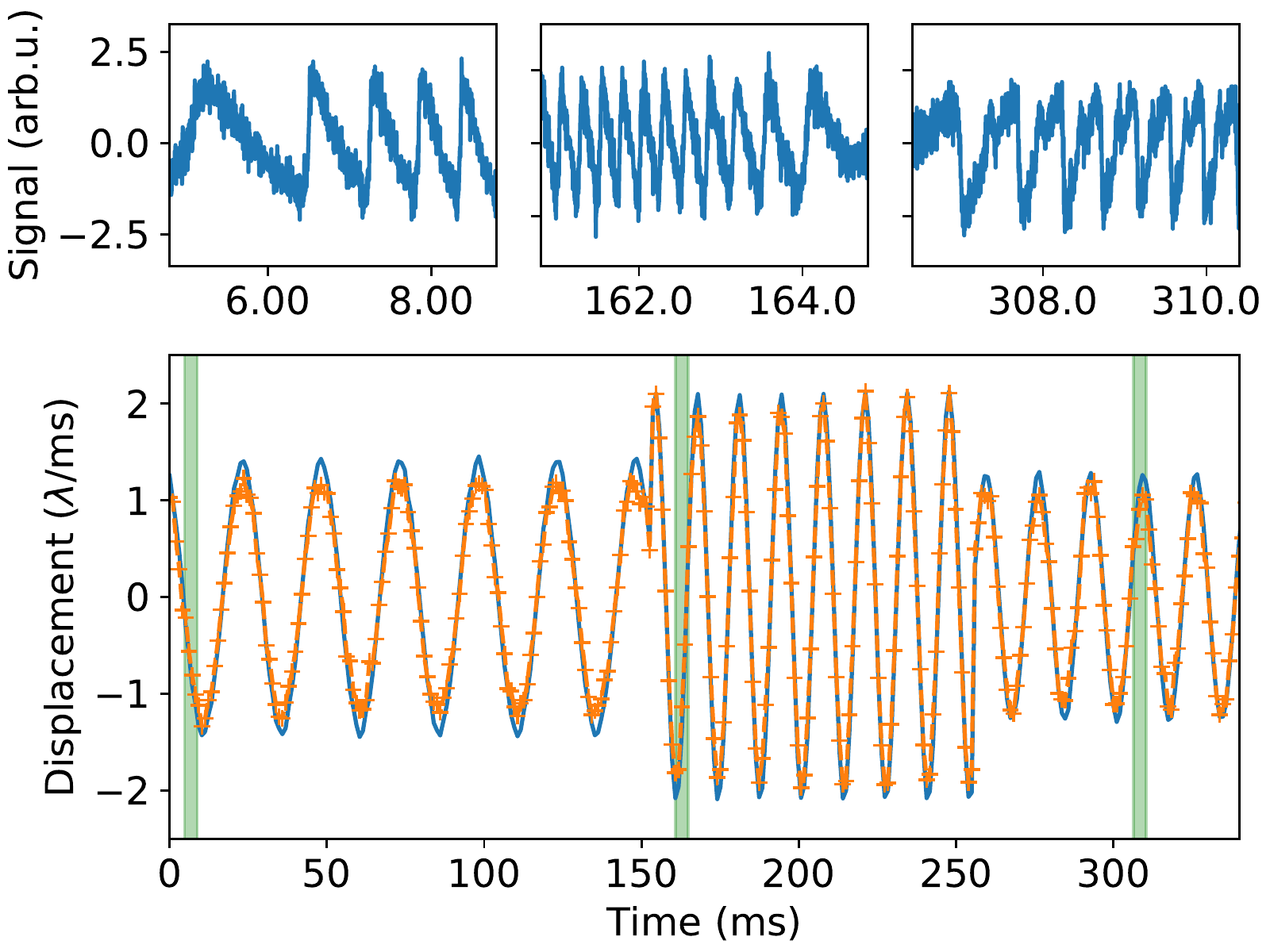}
	\caption{Reconstruction of periodic displacements similar to those used to train the network. Top row: interferometric signal corresponding to three different alignment configurations. Bottom row: measured displacement signal (blue continuous line) and trajectory predicted by the network. The orange crosses are predicted by the model and the dashed line is a simple cubic interpolation. The shaded areas correspond to the interferometric signals shown on the top row. \label{fig:training}}
\end{figure}

\subsection{Periodic signals, known alignment conditions}

We show on \ffig{fig:training} how the network reconstructs examples of periodic traces after training. This exact sequence has not been used during training but these alignment conditions were used during training and sequences with identical frequencies and amplitude were used during training.

On the top row we show three examples of interferometric signal which correspond to three of the alignment conditions used during the training of the network and three different displacement frequencies. On the bottom row, we show the displacement per time unit of the target, as it can be measured from the voltage at the edges of the speaker (blue continuous line). Independently of that voltage measurement, we use the trained neural network to infer the displacement from the self-mixing signal. This is the orange dashed line, which is almost perfectly superimposed to the actual displacement measured from the voltage at the speaker's ends. This almost perfect reconstruction is not very surprising since, even if the network had not seen this exact piece of time trace during training, it has seen periodic signals at these frequencies, these amplitudes and in these exact alignment conditions. That is however a confirmation that the training of the network has worked to an excellent accuracy and \textit{under different alignment conditions}.

\subsection{Aperiodic signals, unknown alignments}
\label{subsec:unknown}

\begin{figure}
	\centering
	\includegraphics[width=0.6\textwidth]{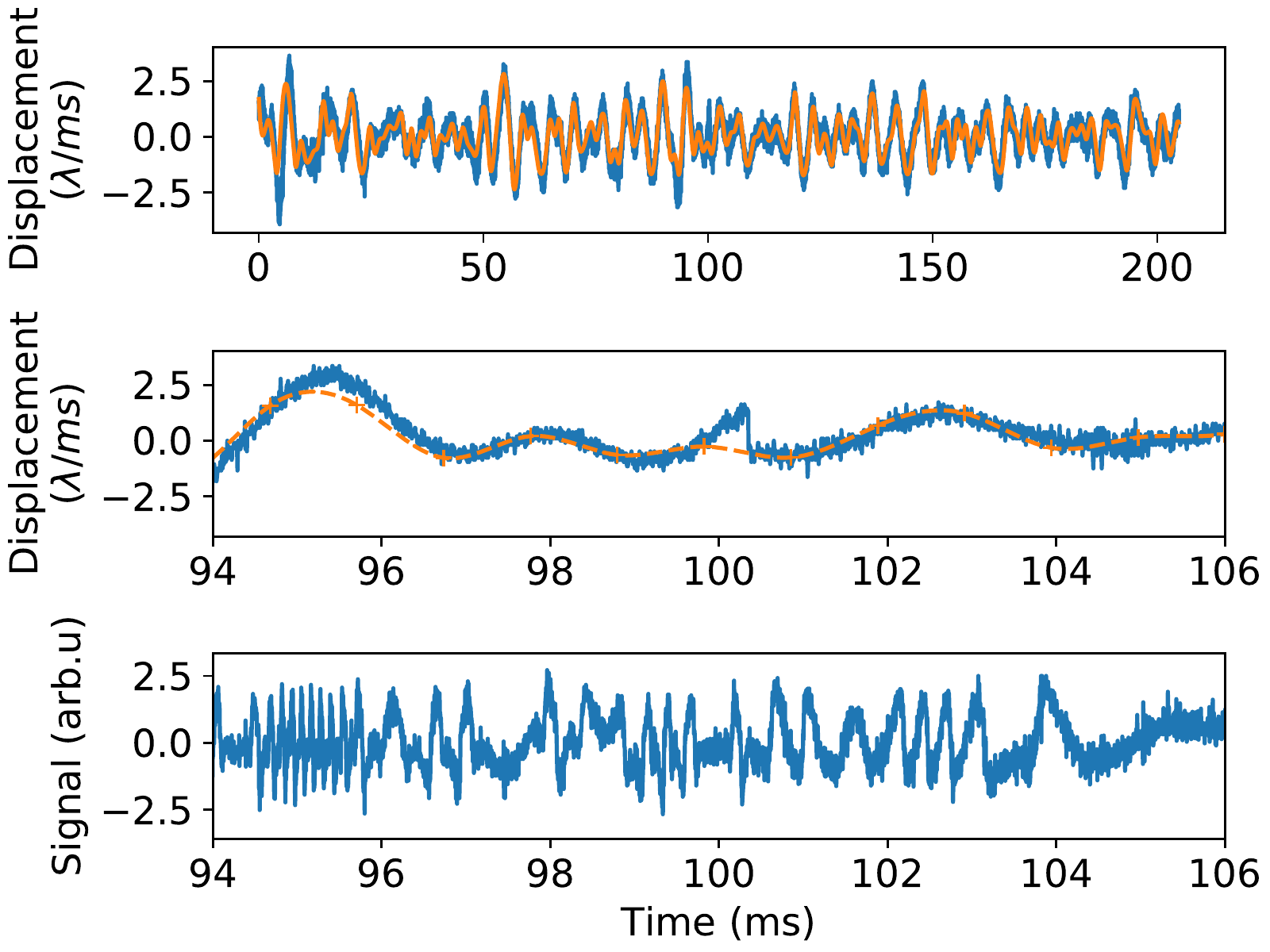}
	\caption{Reconstruction of unknown and complex displacements in alignment situations which are not in the training set. Top: the blue line is the displacement measured from the speaker voltage, the orange line is the model's prediction. Middle row: zoom around the central area of the top row. The discontinuity close to 100~ms is where we numerically connected the two time traces (see text). Bottom row: interferometric signal corresponding to the middle row. \label{fig:unknown}}
\end{figure}

We check the capacity of the statistical model to adapt to unseen situations by preparing a completely different displacement pattern. This pattern is obtained by applying a fifth order butterworth band-pass filter between 5 and 100~Hz to a delta-correlated gaussian random noise. This pattern is sent to the speaker in two different alignment conditions, none of them corresponding the the situations used during training. In one of the two situations, the interferometric signal shows a double-peak structure. The two interferometric signals are then concatenated into a single time series and we use the model to reconstruct the displacement of the target corresponding to this concatenated time series. The results are shown on \ffig{fig:unknown}.

As can be immediately appreciated, the reconstruction is excellent, the prediction matching almost perfectly the independently measured displacement. Of course the discontinuity close to 100~ms, where the two measurements are artificially concatenated, cannot be predicted by the network since it is absent in the measured interferometric signal. We just choose to emphasize this region as it shows that the prediction is essentially insensitive to the alignment conditions, which change abruptly in the middle of the trace. As is evident from the lower panel of \ffig{fig:unknown}, reconstructing a trajectory from this interferometric signal would be very difficult due to the presence of noise and very widely varying fringe shapes and repetition rates. 

From the above, one concludes that the model is able to generalize from its learning set to provide an accurate reconstruction of the displacement in unseen alignment conditions and for very complex time series, much more difficult to analyze than the simple periodic time traces used during the training phase.

\begin{figure}
	\centering
	\includegraphics[width=0.6\textwidth]{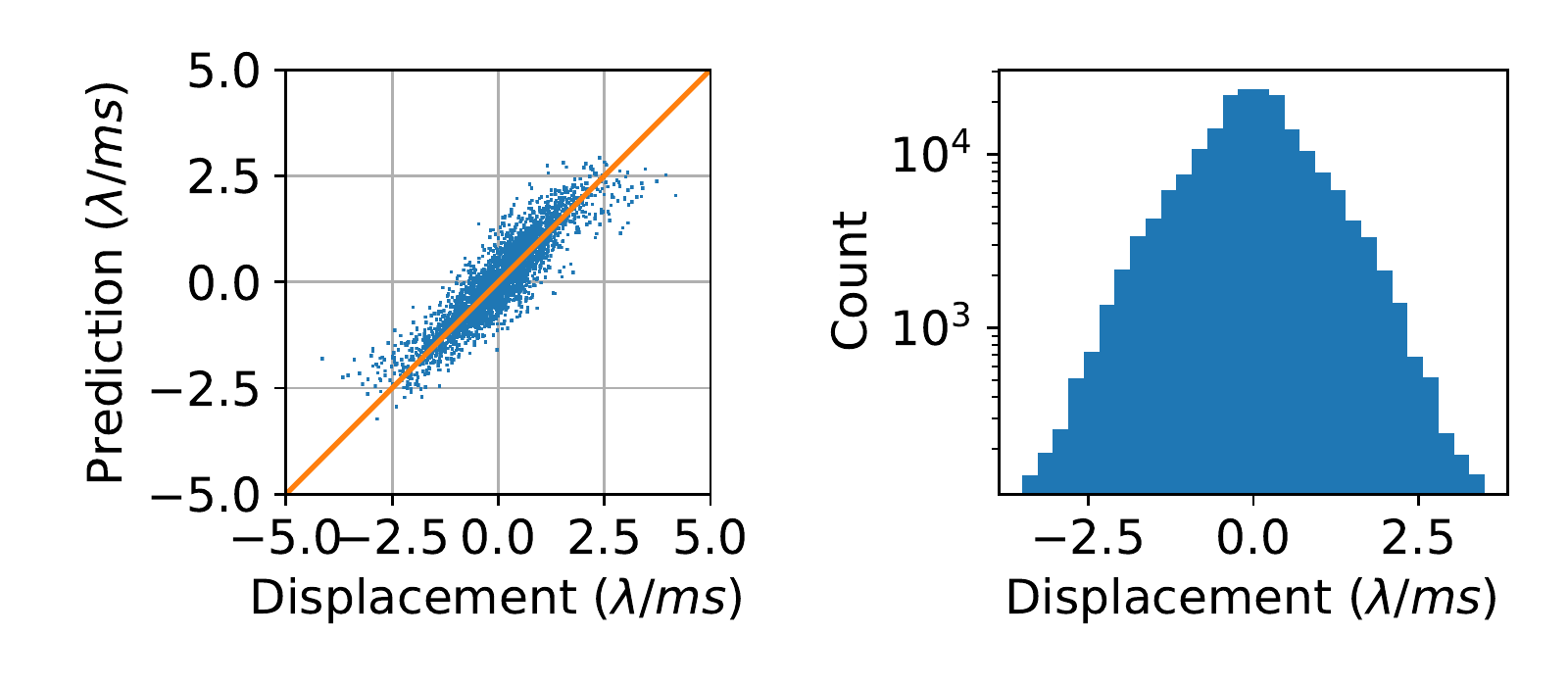}
	\caption{The accuracy of prediction on unknown samples (at a given sampling rate) is conditioned by the training set. Left: displacement inferred by the model as a function of the true displacement. The correlation is excellent but the model slightly underestimates the larger displacements (about 2.5~$\lambda$/ms). Right: histograms of the distribution of displacements in the training sets. The larger displacements (about 2.5~$\lambda$/ms) are very under-represented in the training set.
	\label{fig:correlate}}
\end{figure}

To better appreciate the accuracy of the inference, one can plot the predicted displacement as a function of the actual displacement as shown on \ffig{fig:correlate}. A perfect reconstruction would be the one shown by the orange line where the prediction is exactly equal to the truth. We can quantify the reconstruction quality by the Pearson's correlation coefficient between the reconstruction and the ground truth which is here 0.90 and the absolute standard error which is here $0.30\lambda/ms$. Specifically, one can notice that the prediction is less good for the largest absolute values of displacement. This can be related to the statistical properties of the training set as shown on the right panel of \ffig{fig:correlate}. Here one can appreciate that absolute values of displacement larger than 2.5$\lambda/ms$ have been seen by the network during training only a few hundreds of times, while smaller displacements are much more frequent in our training set. Thus, the large displacements are very under-represented in the training set. This results in a lower precision of the reconstruction for larger displacements, which can also be appreciated on the top panel of \ffig{fig:unknown} where the largest displacements are in general under estimated.

\subsection{Noise sensitivity}

\begin{figure}
	\centering
	\includegraphics[width=0.6\textwidth]{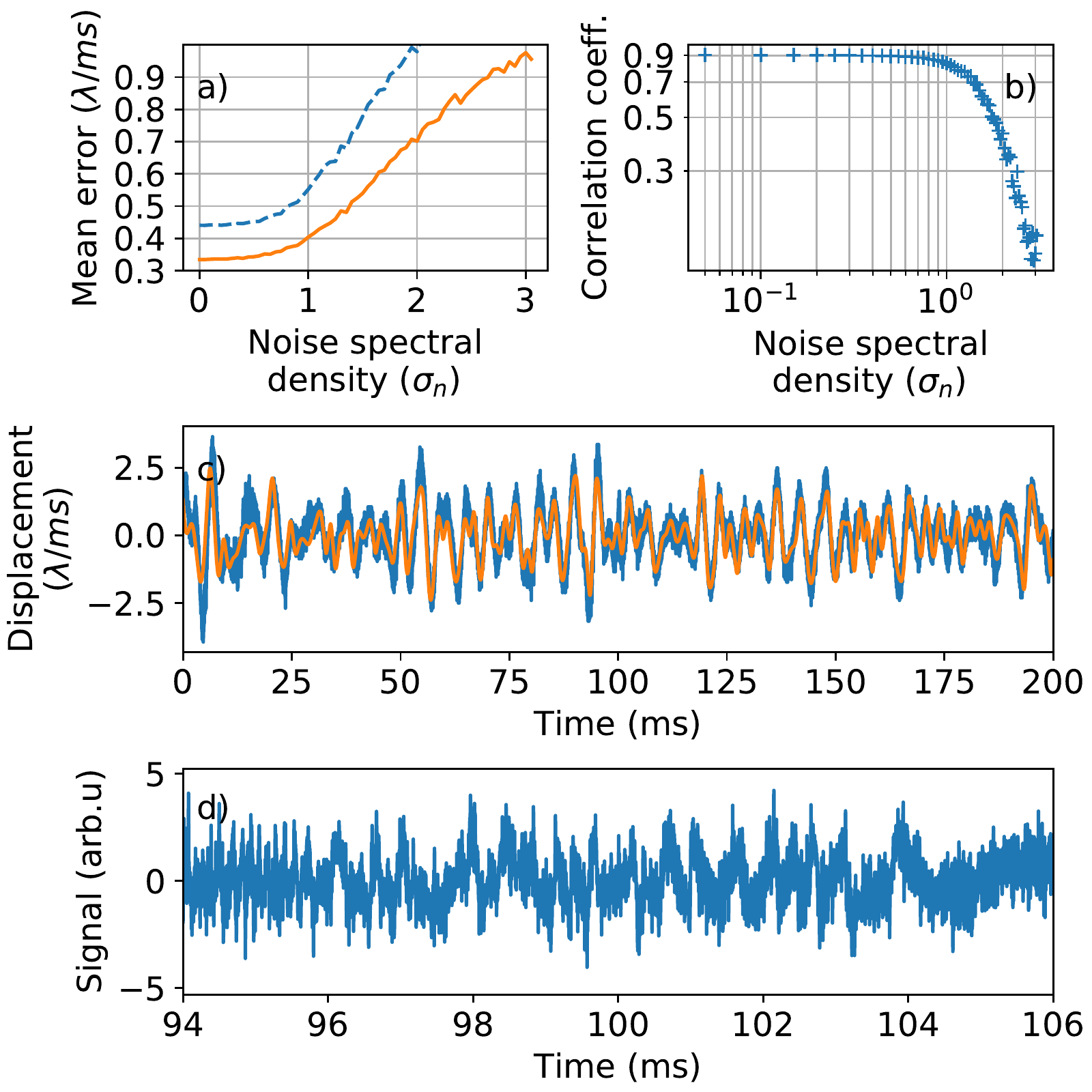}
	\caption{Robustness against detection noise. a) Mean absolute (orange continuous) and RMS (blue dashed) error of the reconstruction as function of added noise in units of the interferometric signal's standard deviation $\sigma_n$. b) Pearson's correlation coefficient between the prediction and the signal as function of added noise. c) Example of prediction (orange line) and true displacement (blue line) for an added noise power density $\sigma_n = 0.7\times\sigma_n$ where $\sigma_n$ is the interferometric signal's standard deviation. d) Interferometric signal in the same situation, zoom over the 94-106~ms region (the same signal as \ffig{fig:unknown}, with added noise).
	\label{fig:noise}}
\end{figure}

One of the difficulties in reconstructing the displacement from the self mixing signal also comes from the fact that simple Fourier filtering is often not very efficient at separating the detection noise from the interferometric signal (although neural networks have been proposed to alleviate this issue \cite{ahmed2019self}). Here we check that the statistical model is very robust to the addition of noise on top of the interferometric signal. To assess this robustness, we use the model to reconstruct the displacement corresponding to the complex interferometric signal described in \ref{subsec:unknown} after adding to this signal a gaussian white noise. As we show on \ffig{fig:noise}, the model predictions are extremely robust. Since the added noise is $\delta-$~correlated, its standard deviation $\sigma_n$ is a measure of its power density. Here, we normalize the interferometric signal itself in absence of added noise to its standard deviation $\sigma_s$ so that $\sigma_s = 1$. We then vary the standard deviation of the added noise $\sigma_n$ between 0 and 3 times $\sigma_s$. On \ffig{fig:noise}a), we observe that both the root mean squared and the absolute error remain very low up to $\sigma_n=1$ where it grows significantly. On \ffig{fig:noise}b), we plot the correlation coefficient between the reconstructed displacement and the independently measured displacement. As for the error, the correlation coefficient indicates an excellent reconstruction of the displacement up to approximately $\sigma_n=1$. We show the prediction and the noisy interferometric signal for $\sigma_n=0.7$ on \ffig{fig:noise} c) and d) respectively. As can be easily observed, the interferometric signal would be rather difficult to process by the usual means and the stochastic model provides a very useful reconstruction.

\subsection{Unknown experiment}
\label{subsec:newexp}

The analysis above has shown that the statistical model is able to reconstruct the displacement from the self-mixing signal in a broad range of unknown conditions. However, all of the above was realized on a single experimental setup. Contrary to a physical model, which is constructed to capture only the universal features of an experiment, an empirically constructed statistical model such as the neural network we use may capture also non-universal and system-specific features. Thus, it is interesting to check what the model can predict on the basis of a \textit{different} experiment, based on the same principle. To address this question, we prepare an "almost-twin" experiment, based on the same self mixing interferometry principle shown in \ffig{fig:setup} but featuring a different laser (HL6323MG, $\lambda=639~nm$, driven at $I=75~mA$ for a threshold current $I_{th}=45~mA$), a different speaker (with a different range of linear response), a different voltage amplifier for the acquisition of the laser diode voltage etc. Although this experiment is in principle the same, it differs in many of the details which should not be relevant to the physics, yet carry a significant risk of distortion of the interferometric signal as compared to the one used in training set. 

\begin{figure}
	\centering
	\includegraphics[width=0.6\textwidth]{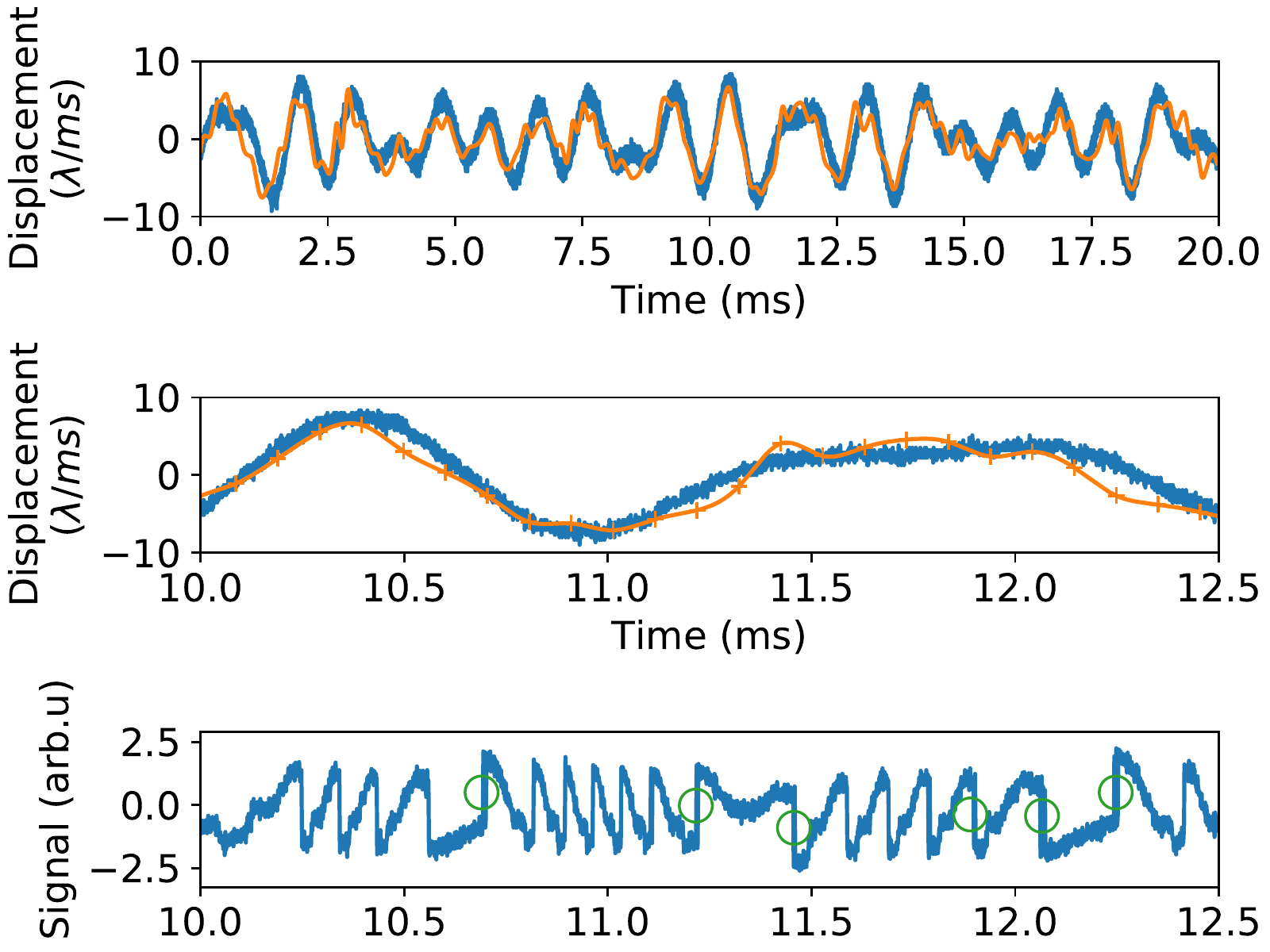}
	\caption{Reconstruction of the displacement on an unknown experiment. Top: comparison between the displacement predicted by the model (orange line) and the actual displacement (blue line). There are no free parameters. Middle: zoom on one specific region. Bottom: self mixing signal corresponding to middle panel. The green circles show where jumps between bistable states occur.
	\label{fig:newexp}}
\end{figure}

To check the model's ability to analyze this new self mixing experiment, we prepare a new displacement time series consisting of the sum of four different frequencies $S = \sum_{i=1}^{4} \sin{2\pi f_i t}$ with $f_i=425, 718, 808, 1076~Hz$. This time series is therefore in a very different (much higher) frequency band with respect to the training experiment. In order to provide the model with comparable input data, the acquisition of the self-mixing signal is performed at a ten times faster rate than in the training experiment (2.5~MHz). The displacement per time unit of the speaker is, as in the previous experiment, measured as a voltage at the edges of the speaker. The comparison between the displacement estimated from the speaker's voltage and the displacement reconstructed from the interferometric signal is shown on \ffig{fig:newexp}.

The agreement between the prediction and the measurement is strikingly good, especially taking into account that no free parameter exist: The model trained on experiment 1 can immediately be used to infer displacements in units of $\lambda/dt$ in experiment 2. 

It is important to underline once more the robustness of this process with respect to specific experimental conditions. For instance, in this experiment, the interferometric signal shows clear signs of bistability between external cavity modes in forms of very fast jumps between states (green circles on bottom panel of \ffig{fig:newexp}). These features are absent from the training set. Here one sees that the model essentially filters them out automatically. Besides this, it is also worth noting that, as compared to \ffig{fig:unknown} for instance, the displacements and time scales are very different. This shows that, provided an adequate sampling rate is chosen, the model can work at much higher frequency than the band it was trained in and for much larger displacements per time unit. This feature is not unexpected since the network knows only about displacement "per $256\times dt$", without reference to the exact value of $dt$. Therefore, with adequate sampling, the measurement range can be tremendously extended with respect to the training range. This feature is extremely useful since it allows training in an easily accessible range (displacement and frequency band) and prediction in very different range for more demanding applications.

\section{Discussion}

The results above clearly show that a convolutional neural network is a useful tool in the reconstruction of a target's displacement, very robust to unknown displacement signal shapes, alignment conditions, electronic noise and even whole setups. We have also verified that a network trained in the 10-100~Hz frequency band can also meaningfully reconstruct displacements including frequencies of hundreds of Hz, provided the measurement sampling rate is adapted. Thus, the neural network strength lies less in the absolute precision that it allows than in its robustness against detailed experimental conditions and versatility across the "sub-wavelength/analog" and "beyond wavelength/digital" classification \cite{donati2018overview} for arbitrarily complex waveforms.

One natural question which arises when preparing a neural network is that of model capacity \cite{goodfellow2016deep}. A network which does not possess enough cells or layers may be unable to take into account all the complexity of the task. On the other hand, a network with a \textit{very large} number of cells and layers will sooner or later learn features of the experiment which should not be significant (for instance, all the details about an amplifier used in the setup). This prevents the network from generalizing, \textit{ie} accurately predicting unknown data. This is in principle dealt with during the training phase \cite{goodfellow2016deep} but it is only when the network processes fully new data that this issue can be totally ruled out. Here this issue has been taken care of by predicting arbitrarily complex trajectories and also by using two different setups. In fact, the imperfect reconstruction in the case of the unknown experiment is most probably due to the model learning some system-specific features of the training experiment. This can be mitigated by a minor retraining of the final layer of the model on the new experiment (a procedure known as "fine tuning" in the deep learning context). We have noticed that a larger network featuring more than $10^5$ coefficients instead of the $5.7 \times 10^4$ used here does not lead to better training and may even lead to worse predictions in the unknown experiment.

The performance of the network is of course strongly related to the training data set which is used. Here we deliberately use only a very limited set of displacements during training in order to very clearly show the generalization phenomenon, the network being able to predict correctly displacement shapes it has never seen before. For real use beyond the proof of concept presented here, more refined training is possible: A training set featuring a more uniform distribution of displacements will provide a more accurate reconstruction of the larger displacements \textit{for a given sampling rate}. As an alternative, simply increasing the sampling rate at the prediction time may also be a sufficient solution to adapt the time series to the operating range of the model as we have shown in \ref{subsec:newexp}. Care must be taken when training the network that the correct operating range of the neural network is set by a displacement per time unit, which includes limitations in terms of displacement frequency and amplitude. Translating it in terms of counting fringes, that means that the network will saturate beyond a certain number of fringes during the $256\times dt$ measurement window. In terms of feedback range, here we have used only $C<1$ in training, avoiding too low values of $C$ where the interferometric signal is symmetric. At the prediction phase, the model is robust to $C$ slightly overcoming unity but when multistability becomes strong the information of few wavelengths displacements is lost and the model has no chance to recover it. Similarly, we have checked that when $C$ is so low that the signal is symmetric, the model cannot predict accurately. A full characterization of performance degradation and the use of multichannel measurements to mitigate this issue is beyond the scope of this work.

One particularly interesting avenue to circumvent the limitations of an experimental training set is to train the network on numerically generated data \cite{kliese2014solving}. One drawback is that the network will not learn more than what is in the physical model used for the simulations, which may be hard in complex settings such as multimode lasers \cite{columbo2012self,columbo2014multimode}. However, numerics may provide a way to obtain a training set for which controlled laboratory experiments would be very hard to realize such as hard shocks or high frequency and high amplitude displacements. Experimental data may then be used to refine the training by using different sampling times as described in the previous paragraph.

Once trained, a convolutional neural network can be used in real time since no pre-processing of the data is required and prediction over thousands of interferometric measurements is very fast: As an example, 3 seconds of signal (749056 interferometric data points) are processed in 0.16 seconds on a standard laptop. In addition, after the initial training, a neural network can relatively easily be repurposed by retraining only its final layers even with a very limited set of data. For instance, it would be particularly interesting to assess the performance of the network trained here on self-mixing schemes which include bias current modulation towards some other sensing task such as refractive index measurements. Alternatively, the input layer of the network can also be reworked at minor cost to take into account multichannel measurements and most interesting would probably be to integrate this approach into multimodality imaging systems \cite{brambilla2020versatile}.

\section{Conclusion}

To conclude, we have presented a detailed analysis of how a reasonably simple convolutional neural network can be used to reconstruct the displacement of a target on the basis of self mixing interferometry. We believe that this approach can become one of the many tools which can be used to tailor or enhance self-mixing coherent sensing setups. Far from being limited to displacement measurement and single mode settings, we believe that computer neural networks can become an extremely useful element of many sensing apparatus, especially self-mixing setups. Finally, we stress that there are very few hard rules about the design of neural networks and this design is in itself often an area of research. The architecture we use here is essentially a simple starting point and many refinements are possible. More specifically, one of the most immediate extensions of the work exposed above is to train a network which can provide an estimation of the accuracy of the reconstruction. This can be achieved by adding a calibrated regression stage on top of the convolutional base prepared here \cite{kuleshov2018accurate}. Other extensions may include more complex network topologies, perhaps mixing convolutional and recurrent layers or including skip connections.

\appendix
\section{Network details}

The key elements of the network are shown on table \ref{tab:network}. We refer the reader to deep learning fundamentals for background information \cite{goodfellow2016deep}. The total number of parameters (57 153) is much smaller than the number of time series segments in the data set even before augmentation. Networks of identical architecture with more cells per layer did not lead to significant improvements. A dropout layer is used here mostly as "safety net" since the data set is very large anyway which makes overfitting improbable. The training of the network takes about ten minutes on a simple GPU (GeForce GTX 1060) and about four times more on CPU (Intel Xeon 3.8GHz).

\begin{table}
	\centering
\begin{tabular}{ |p{2.5cm}|p{2.5cm}|p{2.5cm}|p{2.5cm}|  }
 \hline
 \multicolumn{4}{|l|}{Network structure: sequential} \\
 \hline
 Layer type & Main\newline hyperparameters &  Trainable\newline parameters & Output shape\\
 \hline
 1D convolutional   &  kernel size: 7\newline filters: 16    & 128&   (250, 16)\\
 Max Pooling &  pool size: 2   & 0   &(125, 16)\\
 1D convolutional & kernel size:7\newline filters: 32 & 3616&  (119, 32)\\
 Max Pooling    & pool size: 2 & 0 &  (59, 32)\\
 1D convolutional&   kernel size: 7\newline filters 64  & 14400 &(53, 64)\\
 Max Pooling    & pool size: 2 & 0 &  (26, 64)\\
 Dropout & 10\%  & 0   & (26, 64)\\
 1D convolutional & kernel size: 7\newline filters: 64  & 28736 &(20,  64)\\
 Max Pooling    & pool size: 2 & 0 &  (10, 64)\\
 Fully connected & units: 16 & 10256 & 16 \\
 Fully connected & units: 1 & 17 & 1\\
 \hline
\end{tabular}
	\caption{Main parameters of the network used in this work. The network is a sequence of 1-dimensional convolutional and dropout layers followed by two fully connected layers for the final regression. The total number of trainable parameters is 57 153.
		\label{tab:network}}
\end{table}

%
%
%
%

\section*{Acknowledgments}
The authors thank Dr.~L.~Columbo, Dr.~M.~Dabbicco and Dr.~F.~Pedaci for helpful discussions.

\section*{Disclosures}

The authors declare no conflicts of interest.



\bibliography{selfmixing}

\begin{thebibliography}{10}
\newcommand{\enquote}[1]{``#1''}

\bibitem{giuliani2002laser}
G.~Giuliani, M.~Norgia, S.~Donati, and T.~Bosch, \enquote{Laser diode
  self-mixing technique for sensing applications,}
  {\protect\JournalTitle{Journal of optics A: Pure and applied optics}}
  \textbf{4}, S283 (2002).

\bibitem{kane2005unlocking}
D.~M. Kane and K.~A. Shore, \emph{Unlocking dynamical diversity: optical
  feedback effects on semiconductor lasers} (John Wiley \& Sons, 2005).

\bibitem{donati2012developing}
S.~Donati, \enquote{Developing self-mixing interferometry for instrumentation
  and measurements,} {\protect\JournalTitle{Laser \& Photonics Reviews}}
  \textbf{6}, 393--417 (2012).

\bibitem{taimre2016laser}
T.~Taimre, M.~Nikoli{\'c}, K.~Bertling, Y.~L. Lim, T.~Bosch, and A.~D.
  Raki{\'c}, \enquote{Laser feedback interferometry: a tutorial on the
  self-mixing effect for coherent sensing,} {\protect\JournalTitle{Advances in
  Optics and Photonics}} \textbf{7}, 570--631 (2015).

\bibitem{li2017laser}
J.~Li, H.~Niu, and Y.~X. Niu, \enquote{Laser feedback interferometry and
  applications: a review,} {\protect\JournalTitle{Optical Engineering}}
  \textbf{56}, 050901 (2017).

\bibitem{donati2018overview}
S.~Donati and M.~Norgia, \enquote{Overview of self-mixing interferometer
  applications to mechanical engineering,} {\protect\JournalTitle{Optical
  Engineering}} \textbf{57}, 051506 (2018).

\bibitem{addy1996effects}
R.~C. Addy, A.~W. Palmer, K.~Thomas, and V.~Grattan, \enquote{Effects of
  external reflector alignment in sensing applications of optical feedback in
  laser diodes,} {\protect\JournalTitle{Journal of lightwave technology}}
  \textbf{14}, 2672--2676 (1996).

\bibitem{lv2005effect}
L.~Lv, H.~Gui, J.~Xie, T.~Zhao, X.~Chen, A.~Wang, F.~Li, D.~He, J.~Xu, and
  H.~Ming, \enquote{Effect of external cavity length on self-mixing signals in
  a multilongitudinal-mode fabry--perot laser diode,}
  {\protect\JournalTitle{Applied optics}} \textbf{44}, 568--571 (2005).

\bibitem{norgia2001interferometric}
M.~Norgia, S.~Donati, and D.~D'Alessandro, \enquote{Interferometric
  measurements of displacement on a diffusing target by a speckle tracking
  technique,} {\protect\JournalTitle{IEEE Journal of quantum electronics}}
  \textbf{37}, 800--806 (2001).

\bibitem{zabit2010adaptive}
U.~Zabit, R.~Atashkhooei, T.~Bosch, S.~Royo, F.~Bony, and A.~Rakic,
  \enquote{Adaptive self-mixing vibrometer based on a liquid lens,}
  {\protect\JournalTitle{Optics letters}} \textbf{35}, 1278--1280 (2010).

\bibitem{bernal2014robust}
O.~D. Bernal, U.~Zabit, and T.~M. Bosch, \enquote{Robust method of
  stabilization of optical feedback regime by using adaptive optics for a
  self-mixing micro-interferometer laser displacement sensor,}
  {\protect\JournalTitle{IEEE Journal of Selected Topics in Quantum
  Electronics}} \textbf{21}, 336--343 (2014).

\bibitem{arriaga2014speckle}
A.~L. Arriaga, F.~Bony, and T.~Bosch, \enquote{Speckle-insensitive fringe
  detection method based on hilbert transform for self-mixing interferometry,}
  {\protect\JournalTitle{Applied optics}} \textbf{53}, 6954--6962 (2014).

\bibitem{usman2019detection}
M.~Usman, U.~Zabit, O.~D. Bernal, G.~Raja, and T.~Bosch, \enquote{Detection of
  multimodal fringes for self-mixing-based vibration measurement,}
  {\protect\JournalTitle{IEEE Transactions on Instrumentation and Measurement}}
  \textbf{69}, 258--267 (2019).

\bibitem{siddiqui2017_all}
A.~A. Siddiqui, U.~Zabit, O.~D. Bernal, G.~Raja, and T.~Bosch, \enquote{All
  {Analog} {Processing} of {Speckle} {Affected} {Self}-{Mixing}
  {Interferometric} {Signals},} {\protect\JournalTitle{IEEE Sensors Journal}}
  \textbf{17}, 5892--5899 (2017).

\bibitem{goodfellow2016deep}
I.~Goodfellow, Y.~Bengio, A.~Courville, and Y.~Bengio, \emph{Deep learning},
  vol.~1 (MIT press Cambridge, 2016).

\bibitem{li2019deep}
H.~Li, C.~Zhang, N.~Song, and H.~Li, \enquote{Deep learning-based interference
  fringes detection using convolutional neural network,}
  {\protect\JournalTitle{IEEE Photonics Journal}} \textbf{11}, 1--14 (2019).

\bibitem{reyes2019deep}
A.~Reyes-Figueroa and M.~Rivera, \enquote{Deep neural network for fringe
  pattern filtering and normalisation,} {\protect\JournalTitle{arXiv preprint
  arXiv:1906.06224}}  (2019).

\bibitem{kou2020fringe}
K.~Kou, C.~Wang, T.~Lian, and J.~Weng, \enquote{Fringe slope discrimination in
  laser self-mixing interferometry using artificial neural network,}
  {\protect\JournalTitle{Optics \& Laser Technology}} \textbf{132}, 106499
  (2020).

\bibitem{wei2007pre}
L.~Wei, J.~Chicharo, Y.~Yu, and J.~Xi, \enquote{Pre-processing of signals
  observed from laser diode self-mixing intereferometries using neural
  networks,} in \emph{2007 IEEE International Symposium on Intelligent Signal
  Processing,}  (IEEE, 2007), pp. 1--5.

\bibitem{ahmed2019self}
I.~Ahmed, U.~Zabit, and A.~Salman, \enquote{Self-mixing interferometric signal
  enhancement using generative adversarial network for laser metric sensing
  applications,} {\protect\JournalTitle{IEEE Access}} \textbf{7},
  174641--174650 (2019).

\bibitem{wang2020real}
X.-l. Wang, L.-p. L{\"u}, L.~Hu, and W.-c. Huang, \enquote{Real-time human
  blood pressure measurement based on laser self-mixing interferometry with
  extreme learning machine,} {\protect\JournalTitle{Optoelectronics Letters}}
  \textbf{16}, 467--470 (2020).

\bibitem{chollet2015keras}
F.~Chollet \emph{et~al.}, \enquote{keras,}  (2015).

\bibitem{hornik1989multilayer}
K.~Hornik, M.~Stinchcombe, H.~White \emph{et~al.}, \enquote{Multilayer
  feedforward networks are universal approximators.}
  {\protect\JournalTitle{Neural networks}} \textbf{2}, 359--366 (1989).

\bibitem{kliese2014solving}
R.~Kliese, T.~Taimre, A.~A.~A. Bakar, Y.~L. Lim, K.~Bertling, M.~Nikoli{\'c},
  J.~Perchoux, T.~Bosch, and A.~D. Raki{\'c}, \enquote{Solving self-mixing
  equations for arbitrary feedback levels: a concise algorithm,}
  {\protect\JournalTitle{Applied optics}} \textbf{53}, 3723--3736 (2014).

\bibitem{columbo2012self}
L.~Columbo, M.~Brambilla, M.~Dabbicco, and G.~Scamarcio, \enquote{Self-mixing
  in multi-transverse mode semiconductor lasers: model and potential
  application to multi-parametric sensing,} {\protect\JournalTitle{Optics
  express}} \textbf{20}, 6286--6305 (2012).

\bibitem{columbo2014multimode}
L.~Columbo and M.~Brambilla, \enquote{Multimode regimes in quantum cascade
  lasers with optical feedback,} {\protect\JournalTitle{Optics Express}}
  \textbf{22}, 10105--10118 (2014).

\bibitem{brambilla2020versatile}
M.~Brambilla, L.~L. Columbo, M.~Dabbicco, F.~De~Lucia, F.~P. Mezzapesa, and
  G.~Scamarcio, \enquote{Versatile multimodality imaging system based on
  detectorless and scanless optical feedback interferometry—a retrospective
  overview for a prospective vision,} {\protect\JournalTitle{Sensors}}
  \textbf{20}, 5930 (2020).

\bibitem{kuleshov2018accurate}
V.~Kuleshov, N.~Fenner, and S.~Ermon, \enquote{Accurate uncertainties for deep
  learning using calibrated regression,} in \emph{Proceedings of the 35th
  International Conference on Machine Learning,}  vol.~80 of \emph{Proceedings
  of Machine Learning Research} J.~Dy and A.~Krause, eds. (PMLR,
  Stockholmsmässan, Stockholm Sweden, 2018), pp. 2796--2804.

\end{thebibliography}

\end{document}